\documentclass[unnumsec,webpdf, modern, large]{mam-authoring-template}%

\usepackage{setspace}

\usepackage[mathlines]{lineno}
\usepackage{booktabs}
\graphicspath{{Fig/}}

\theoremstyle{thmstyleone}%
\theoremstyle{thmstyletwo}%
\theoremstyle{thmstylethree}%

\begin{document}
\onecolumn

\journaltitle{Microscopy and Microanalysis}
\DOI{DOI HERE}
\copyrightyear{2024}
\pubyear{2014}
\access{Advance Access Publication Date: Day Month Year}
\appnotes{Original Article}

\firstpage{1}



\title[On Transmission Function Amplitude and Phase Recovery in Multislice Electron Ptychography 
]{On Transmission Function Amplitude and Phase Recovery in Multislice Electron Ptychography 
}

\author{Bridget R.~Denzer\ORCID{0000-0003-4054-1584}}
\author{Colin Gilgenbach\ORCID{0000-0003-0847-8319}}
\author[$\ast$]{James M.~LeBeau\ORCID{0000-0002-7726-3533}}

\authormark{Bridget R.~Denzer et al.}

\address[1]{\orgdiv{Department of Materials Science \& Engineering}, \orgname{Massachusetts Institute of Technology}, \orgaddress{\street{Cambridge}, \postcode{02139}, \state{MA}, \country{USA}}}

\corresp[$\ast$]{Corresponding author. \href{email:email-id.com}{lebeau@mit.edu}}

\received{Date}{0}{Year}
\revised{Date}{0}{Year}
\accepted{Date}{0}{Year}




\abstract{Multislice electron ptychography commonly accounts for inelastic scattering by including an absorptive object potential in the reconstruction forward model that attenuates the elastic signal. However, it remains unclear how the reconstruction is quantitatively impacted by thermal scattering. Here, we use the quantum excitation of phonons (QEP) formalism, which explicitly separates elastic and inelastic (thermal diffuse) scattering within a single multislice propagation, to simulate four-dimensional scanning transmission electron microscopy (4D STEM) datasets of PbTiO$_3$ and SrTiO$_3$ and reconstruct their phase and amplitude. We find that reconstructions of the QEP total and elastic-only datasets are nearly indistinguishable across sample thicknesses from 11 to 40 nm and across atomic species, demonstrating that the reconstruction is largely insensitive to incoherent thermal diffuse scattering within the collected angular range. Reconstructions from multislice phase-only simulations further confirm that appreciable amplitude does not arise from multiple elastic scattering, but instead reflects attenuation of the coherent elastic signal due to thermal scattering. Further comparison of the QEP elastic-only reconstruction with absorptive multislice simulations reveals substantial amplitude contrast deviations for heavy Pb columns (up to 17\%), arising from the approximations used to evaluate the absorptive potential for the 4D STEM simulation. These results thus indicate that reconstruction phase and amplitude accuracy are not significantly limited by the absorptive potential forward model, even in the presence of thermal diffuse scattering.}



\keywords{ptychography, amplitude, quantitative analysis, 4D STEM, thermal diffuse scattering}


\maketitle
\section{Introduction}
\doublespacing



Iterative multislice electron ptychography reconstructs a stack of transmission functions and the incident wavefunction from 4D STEM diffraction patterns, thereby accounting for dynamical scattering \citep{chen_electron_2021, jiang_electron_2018, chen_mixed-state_2020}. Implementations to date have largely relied on approximations that account for incoherent scattering, e.g., thermal diffuse scattering (TDS), using a complex electrostatic potential \citep{gilgenbach_phaser_2026, chen_electron_2021, jiang_electron_2018, odstrcil_iterative_2018, wakonig_ptychoshelves_2020, lee_ptyrad_2025}. Although thermal scattering models -- the absorptive-potential, frozen-phonon, and quantum excitation of phonons methods -- have been extensively studied in conventional microscopy \citep{alania_frozen_2018, peters_fast_2021, terzoudis-lumsden_quantitative_2025, martin_model_2009, van_dyck_is_2009}, it remains less clear how using the absorptive potential within the reconstruction forward model limits object recovery, and how this limitation depends on thickness and atomic number, which is essential for quantitative analysis.

While recent approaches such as multi-object-mode reconstructions
\citep{gladyshev_electron_2023, diederichs_exact_2024, herdegen_thermal_2024} aim to account for thermal scattering more rigorously, absorptive-potential forward models are likely to remain widely used for their computational efficiency. In this model, the complex projected potential, $V_p = V_r + iV_i$, comprises a real component, $V_r$, that models elastic, coherent scattering and an imaginary component, $V_i$, that accounts for inelastic, incoherent scattering. As the transmission function, $t$, of a slice is given by $t = e^{i\sigma V_p}$, the phase shift introduced by elastic scattering is $\sigma V_r$, while the amplitude is $e^{-\sigma V_i}$. Thus, inelastic scattering accounted for by $V_i$ leads to attenuation of the outgoing intensity. Consequently, $V_i$ is referred to as an absorptive potential and can account for phonon scattering, higher-energy losses, and other sources of incoherence originating from the sample \citep{peng_debyewaller_1996, weickenmeier_computation_1991, peng_electron_1999, yoshioka_effect_1957, hall_effect_1965}. 

Although multislice calculations utilizing a complex electrostatic potential are computationally efficient, this approach neglects multiple scattering of inelastically scattered electrons, which is needed to reproduce certain diffraction features such as Kikuchi bands \citep{loane_thermal_1991} and thermal diffuse scattering \citep{lebeau_quantitative_2008}.  The consequences of this simplification for the reconstructed object, however, cannot be assumed a priori. Unlike in the 4D STEM simulations themselves, the effects of these thermal-scattering assumptions on the ptychographic reconstructions are not always straightforward to predict, since the nonlinear, nonconvex nature of the reconstruction algorithm can obscure their impact. This is particularly true for the transmission function amplitude, which is dominated by thermal contributions \citep{wang_thermal_1993} and is therefore likely to be highly sensitive to the choice of thermal scattering model. Moreover, the strong atomic number ($Z$) contrast of the reconstructed amplitude, similar to that of high-angle annular dark-field (HAADF) STEM, offers significant potential for quantitative compositional analysis \citep{denzer_optimizing_2025}, but realizing this potential requires a deeper understanding of the amplitude signal and its dependence on the thermal scattering model.

To simulate the effects of multiple thermal scattering, the frozen-phonon and quantum excitation of phonons (QEP) methods are commonly used, both of which account for multiple thermal scattering at significantly greater computational cost \citep{kirkland_advanced_2010, loane_thermal_1991}. Of these methods, QEP considers many thermal displacements at every slice within a single propagation, allowing the electron to undergo multiple, depth-correlated inelastic scattering events as it traverses the specimen, and further enabling separation of elastically and inelastically scattered electrons in the final diffracted intensities \citep{allen_modelling_2015}. QEP is thus  particularly useful to quantify the effects of multiple thermal scattering on ptychography reconstructions.

Here, we investigate how thermal scattering affects multislice electron ptychographic reconstructions by fixing the reconstruction model to a complex, absorptive potential while varying the treatment of thermal scattering in 4D STEM simulations of PbTiO$_3$ (PTO) and SrTiO$_3$ (STO), for sample thicknesses up to 40 nm. Using the QEP formalism to explicitly separate the elastic and inelastic components within a single propagation, we assess the effect of each on the reconstruction independently, including that of the characteristic Kikuchi-band intensity \citep{herdegen_thermal_2024}. We first show that the reconstructed phase and amplitude are nearly insensitive to incoherent thermal diffuse scattering, with reconstructions of the QEP total and elastic-only datasets in excellent agreement across all thicknesses and species considered. Reconstructions of a multislice phase-only object then show that appreciable amplitude contrast appears only when attenuation of the coherent elastic signal is included in the simulation, identifying this attenuation as the origin of the reconstructed amplitude. Comparing reconstructions from the QEP elastic-only and absorptive-potential simulations further reveals substantial amplitude contrast differences at heavy Pb columns that increase with thickness, which we trace to the approximations used to evaluate the absorptive potential at high $Z$. Together, these results show that incoherent thermal diffuse scattering does not limit quantitative phase and amplitude analysis in ptychography, so inexpensive absorptive-potential simulations can largely stand in for full thermal models when assessing ptychographic reconstruction fidelity.

\section{Methods}\label{methods}

\subsection{4D STEM Simulations}\label{4D STEM Simulations}

Multislice 4D STEM simulations of PbTiO$_3$ (PTO) and SrTiO$_3$ (STO) were performed using the MuSTEM simulation package \citep{allen_modelling_2015} to explicitly separate the coherent elastic and inelastic TDS contributions. The supercell comprised 8 $\times$ 8 $\times$ 74 perovskite unit cells ($a = 0.39$ nm) and was oriented along $[001]$, yielding a thickness of 29 nm, with additional simulations spanning thicknesses from 11 to 40 nm. The probe wavefunction and transmission functions were sampled to a bandwidth limit of 645 mrad. The slice thickness in the simulation was 0.195 nm. The probe was formed with 300 keV electrons, a convergence semi-angle ($\alpha$) of 25 mrad, and a defocus of 15 nm (overfocus). Finite source size effects were not included.

Simulations were carried out using either a multislice phase-only potential with unity amplitude, a complex atomic potential accounting for thermal scattering, or the quantum excitation of phonons (QEP) formalism, all simulated using the MuSTEM software. For PTO, the Debye-Waller factors used were B$_{\text{Pb}}$ = 0.687 Å$^2$, B$_{\text{Ti}}$ = 0.337 Å$^2$, and B$_{\text{O}}$ = 0.592 Å$^2$ \citep{nelmes_crystal_1985}, and for STO they were B$_{\text{Sr}}$ = 0.450 Å$^2$, B$_{\text{Ti}}$ = 0.319 Å$^2$, and B$_{\text{O}}$ = 0.418 Å$^2$ \citep{jauch_electron-density_2005}. The phase-only simulations used the thermally-blurred elastic potential only, retaining the real component $V_r$ but excluding the imaginary absorptive term $V_i$. The QEP datasets employed 50 thermal configurations (Einstein approximation) unless otherwise noted.

The 4D STEM dataset used a detector modeled on the electron microscope pixel array detector (EMPAD) \citep{tate_high_2016}. With 128$\times$128 pixels sampled at 0.63 mrad/pixel, the detector spanned $\pm$40 mrad along each axis ($\theta_{x} = \theta_{y} = 40$ mrad at the edge). The probe was scanned over a grid of 74 $\times$ 74 scan positions with a step size of 0.42 Å. Finite electron dose effects were incorporated by adding Poisson noise corresponding to a dose of 1.2 $\times$ 10$^5$ e$^-$/Å$^2$. 

\subsection{Reconstructions and Analysis}\label{reconstructions}

Multislice electron ptychography reconstructions were performed using the \texttt{phaser} Python package with gradient descent \citep{gilgenbach_phaser_2026}. All reconstructions used 6 probe modes and 2 engines, the first running 300 iterations with 2 nm slices and the second 100 iterations with 1 nm slices. In both engines, the object, probe, and position learning rates were 0.01, 0.001, and 0.05, respectively, and L1 and L2 regularizers were applied with a cost of 0.5 each. The following constraints were applied for each iteration: limit probe support (\texttt{max\_angle} = 26.0), clamped object amplitude (\texttt{amplitude} = 1.0), Gaussian layer regularization (\texttt{sigma} = 100, \texttt{weight} = 0.3), object phase Gaussian (\texttt{sigma} = 0.3, \texttt{weight} = 0.3), and object amplitude Gaussian (\texttt{sigma} = 0.3, \texttt{weight} = 0.3). Convergence was assessed using the detector loss; all reconstructions reached changes of less than 0.05\% between consecutive 20-iteration intervals by 100 iterations in the first engine and within 70 iterations in the second.

 
For comparison with the reconstructions, the transmission function was resampled to a real-space pixel size of 25 pm/px to match the reconstruction pixel size and blurred with a Gaussian of 0.3 Å full width half maximum to approximate the object blur introduced during reconstruction and allow comparison of the atom-column line profiles. Both the simulation transmission function and the reconstruction were evaluated over an equivalent thickness of three unit cells (1.2 nm). For the simulation, this corresponded to six slices, summed for phase and multiplied for amplitude. An additional set of reconstructions at 29 nm thicknesses was performed with 2 nm and 1.2 nm slice thicknesses in the first and second engines, respectively, and an average 1.2 nm slice was analyzed (arithmetic mean for phase and geometric mean for amplitude, consistent with the multiplicative amplitude) after removing the top and bottom four slices to exclude vacuum contributions.

Phase ramps were removed from both the resampled transmission function phase and reconstructed phase by fitting a surface to the regions between atom columns and subtracting it from each slice. 
Peak phase and amplitude values, along with atom column positions, were measured by fitting a 2D Gaussian to each column with the LMFIT Python package \citep{newville_lmfit_2025}. Mean values were taken over a circular region (r = 0.75 Å) centered on each fitted position with sub-pixel precision. Unless otherwise noted, amplitude contrast refers to the contrast in the transmission function attenuation (1-amplitude) due to the unity amplitude baseline.

\section{Results and Discussion}\label{results}

\subsection{Sensitivity of the Reconstruction to Thermal Scattering}\

\begin{figure}[htbp]
    \centering
    \includegraphics[width=3.2in]{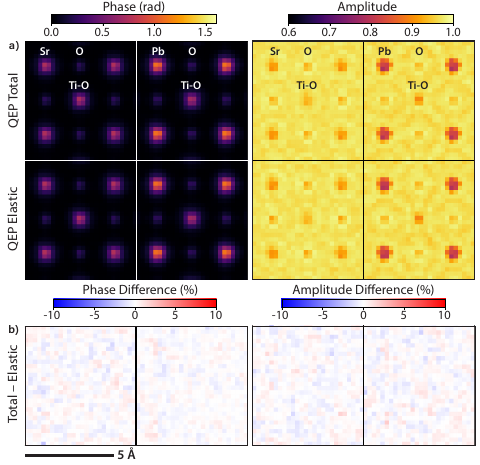}
    \caption{(a) Reconstructed phase and amplitude of 29 nm thick STO (left) and PTO (right) from QEP total and QEP elastic-only datasets. The images represent a 1 nm slice thickness, averaged through the depth after removing the top and bottom four slices. (b) Difference maps between the QEP total and elastic-only reconstructions for STO (left) and PTO (right), normalized to the maximum value from the QEP total image.}
    \label{fig:mustem_ptycho_recons}
\end{figure}

\begin{figure}[htbp]
    \centering
    \includegraphics[width=3.2in]{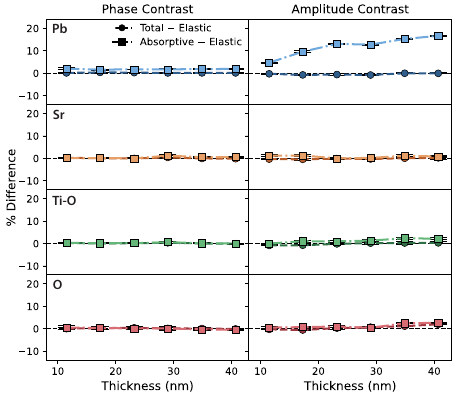}
    \caption{QEP Total -- Elastic and Absorptive -- Elastic percent differences of mean atom column reconstructed phase and amplitude contrast for Pb, Sr, Ti-O, and O atom columns. Mean atom column values are the mean signal over a circular region with radius = 0.75 Å normalized by the area. The Ti-O and O mean values are averaged from the STO and PTO reconstructions. Error bars shown represent the standard deviation from 8 repeat reconstruction runs.}
    \label{fig:ptycho_species_across_thickness}
\end{figure}

Reconstructions from the QEP total (elastic and inelastic) and elastic-only 4D STEM datasets are first compared for a 29 nm thick sample, shown in Figure \ref{fig:mustem_ptycho_recons}a for STO and PTO. In both cases, clear atom-column contrast is recovered in both the reconstructed phase and amplitude, including from the QEP elastic-only dataset. Notably, the phase and amplitude are in excellent agreement with $<2\%$ deviation in mean atom signal for all species between the two cases, with differences at the noise level (Figure \ref{fig:mustem_ptycho_recons}b). Moreover, this excellent agreement persists across the full thickness range studied (up to 40 nm) for both STO and PTO (Figure \ref{fig:ptycho_species_across_thickness}).

Therefore, incoherent thermal scattering, including the characteristic Kikuchi-band intensity, has a negligible effect on the reconstructed phase and amplitude atom-column contrast at these thicknesses and collection angle. Although a full investigation of reconstruction regularization is beyond the scope of this work, additional parameter sweeps preserve the presence and overall trends of the reconstructed atom-column contrast, Figure~S1 in the Supplementary Information, suggesting that the conclusions are robust across a range of reconstruction regularization parameters. This insensitivity is not a matter of detectability. At the dose used here, the diffuse intensity recorded within the collection range lies above the detector noise floor and is measurable above the electron shot noise \citep{tate_high_2016}, so the reconstruction is insensitive to a thermal diffuse contribution that is measurably present in the data. It does, however, reflect the angular range recorded. The detector collects to $\pm$40 mrad, and the higher-angle intensity that is predominantly thermal diffuse in origin falls outside the detector and therefore never enters the reconstruction, so this result should not necessarily be assumed to extend to substantially larger collection angles.

\subsection{Origins of the Reconstructed Phase and Amplitude Contrast}

To further explore the amplitude contrast, reconstructions from the QEP elastic-only datasets are directly compared to those from 4D STEM multislice simulations that use an absorptive potential. Specifically, we simulate two objects: one phase-only, using thermally blurred elastic potentials ($V_p = V_r$), and one that additionally includes thermal absorption ($V_p = V_r + iV_i$). In both cases the transmission function $e^{i\sigma V_p}$ is complex-valued, but only the absorptive potential reduces its modulus below unity at the atomic column positions.

\begin{figure}[htbp]
    \centering
    \includegraphics[width=3.2in]{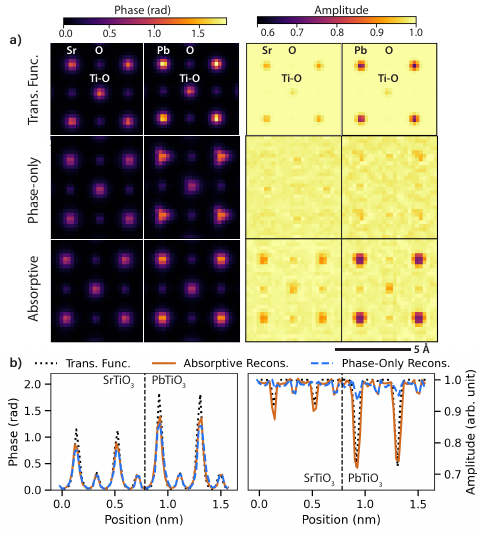}
    \caption{(a) Phase (rad) and amplitude from ptychography reconstructions of an average 1.2 nm slice from a 29 nm thick sample of (left) SrTiO$_3$ and (right) PbTiO$_3$ from the transmission function, reconstruction of a phase-only potential, and reconstruction of an absorptive potential. The corresponding line profiles for STO and PTO are shown in (b). For comparison with the transmission function, the reconstructed amplitudes are background subtracted in (b) and (c).}
    \label{fig:tf_ptycho}
\end{figure}

Reconstructing the phase-only 4D-STEM simulations of 29~nm thick STO and PTO recovers no absorptive component. The reconstructed amplitude deviates from unity only by small, noise-level fluctuations (Figure~\ref{fig:tf_ptycho}). 
Further, the reconstructed phase does exhibit triangular artifacts at the strongly scattering Pb columns while the lower-$Z$ columns remain round. The precise origins are unclear, but may result from the strong scattering at these columns or as an artifact of reconstruction. Thus, a phase-only object does not recover amplitude contrast regardless of thickness or the degree of multiple elastic scattering.  

With an absorptive potential included in the simulation, the reconstruction recovers both phase and amplitude, resolving Pb, Sr, and O columns in both signals and showing marked amplitude contrast between the Sr and Pb sites (Figure~\ref{fig:tf_ptycho}b). Importantly, the attenuation described by the absorptive potential represents the physical loss of intensity from the coherent elastic channel due to thermal scattering. Taken together with the phase-only results, this suggests that reconstructions constrained to a pure or near-pure phase object, such as those that clamp the recovered transmission-function amplitude to remain near unity, do not correctly capture the propagation of the elastic wave.

Next, a depth-averaged reconstruction slice is compared with the corresponding transmission function over the three unit cells spanned by that slice. Because phase and amplitude contrast depend on reconstruction parameters and contrast transfer effects, relative contrast between species is considered rather than absolute values. The reconstructed Pb-to-Sr amplitude contrast (94\%) is approximately twice the reconstructed phase contrast (46\%), an amplitude-to-phase ratio of 2.0. The transmission function gives a comparable ratio of 2.2 (114\% and 51\%, respectively). This is consistent with the stronger $Z$-dependence of the amplitude ($\sim Z^2$) compared with the phase ($\sim Z^{0.67}$), which enhances the separation between the heavier Pb and lighter Sr columns \citep{denzer_optimizing_2025}.

In contrast to the close agreement between the QEP total and elastic-only reconstructions, substantial deviations in amplitude contrast emerge at the heavy Pb columns. The mean Pb amplitude contrast exhibits a relative difference of 13\% at 29 nm thickness, increasing to 17\% at 40 nm (Figure~\ref{fig:ptycho_species_across_thickness}). The corresponding phase difference is much smaller ($\sim$2\%), and all other species differ by less than 1\% in phase and 3\% in amplitude. The QEP elastic and absorptive-potential models therefore do not describe coherent elastic scattering equivalently, and the discrepancy is confined to the most strongly scattering columns. Why the two models diverge for heavy columns, and why the effect grows with thickness, is discussed in the next section.

\subsection{Breakdown of the Absorptive Potential at High $Z$}
\label{sec:amp_mismatch}

\begin{figure}[htbp]
    \centering
    \includegraphics[width=3.2in]{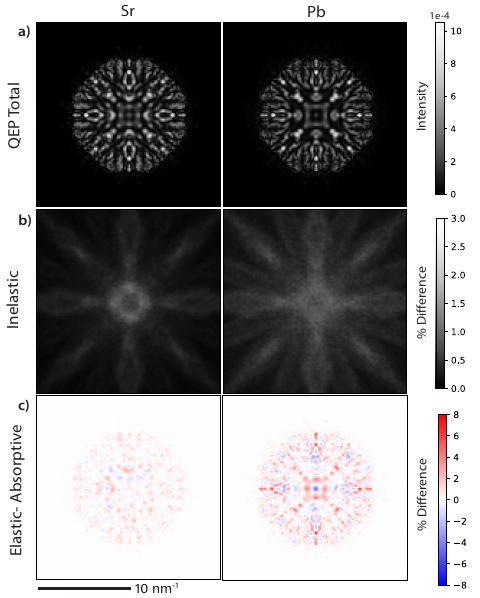}
    \caption{(a) QEP total CBED patterns and (b) QEP inelastic for  Sr (left) and  Pb (right) atom columns for a 29 nm thick sample. (c) Elastic - absorptive CBED difference maps, normalized to the maximum value in the corresponding elastic CBED pattern. The QEP CBED patterns are generated with 200 thermal configurations.}
    \label{fig:mustem_cbed}
\end{figure}

The QEP elastic-only and absorptive-potential models are compared through convergent beam electron diffraction (CBED) patterns from the multislice simulations. Figure~\ref{fig:mustem_cbed}a shows CBED patterns from a QEP total simulation for Sr and Pb atom columns in a 29 nm thick sample. Subtracting the elastic-only pattern from the total isolates the TDS contribution, which appears as diffuse intensity with characteristic Kikuchi bands (Figure~\ref{fig:mustem_cbed}b). The Pb elastic -- absorptive difference instead reveals a non-random redistribution of intensity within the bright-field (BF) disk with a much weaker difference at the Sr columns, Figure~\ref{fig:mustem_cbed}c. This structure sharpens as the number of thermal configurations increases from 50 to 1000, as shown in Figure~S2 in the Supplementary Information, so it reflects a systematic difference between the two models rather than residual statistical noise.

These differences between the QEP elastic and absorptive CBED patterns can be understood from the approximations underlying the absorptive-potential model. While the form of the absorptive potential is physically well founded, the values of $V_i$ are conventionally evaluated as the second-order approximation to the thermally-averaged transmission function, which \citet{anstis_corrections_1996} showed to be reliable for light elements such as Al but substantially in error for heavy elements such as Au, where an additional real correction is also required but commonly omitted. This trend matches the results shown here. Further, reconstructions of the QEP elastic and absorptive simulations agree closely for the weaker-scattering O, Ti-O, and Sr columns, while the Pb columns ($Z = 82$, comparable to Au) differ substantially, and the difference grows with thickness, consistent with the deviation of the absorptive potential approximation with $Z$ and thickness.

\section{Conclusions}

Over the thickness range and with the reconstruction regularization used here, incoherent thermal diffuse scattering has little effect on the phase and amplitude recovered by multislice electron ptychography. Using the QEP formalism to separate the coherent elastic and incoherent contributions, reconstructions of the total and elastic-only datasets agree to within a few percent for all atomic species, despite the characteristic Kikuchi-band intensity present in the total datasets. This insensitivity to thermal diffuse scattering is bounded by the $\pm$40 mrad collection range used here, beyond which the recorded intensity becomes increasingly thermal diffuse in origin. The amplitude contrast instead originates in the attenuation of the coherent elastic signal by thermal scattering. Consistent with this, phase-only simulations, which retain multiple elastic scattering and Debye-Waller blurring but omit the absorptive term, yield no appreciable reconstructed amplitude contrast. Reconstructions of the QEP elastic-only and absorptive-potential simulations nonetheless deviate substantially at the heavy Pb columns, with amplitude contrast differing by up to 17\% over this thickness range while the phase differs by only a few percent. Simulated CBED patterns show a corresponding structured intensity difference at the Pb columns, consistent with the limitations of the conventional absorptive-potential approximation becoming unreliable for high-$Z$ elements. The difference between the two models originates in their treatment of coherent elastic scattering rather than in the diffuse intensity.

The absorptive-potential forward model consequently remains accurate to within a few percent for quantitative phase analysis at low computational cost, whereas quantitative interpretation of the reconstructed amplitude at heavy atom columns calls for a fuller treatment such as QEP or higher-order corrections to the absorptive potentials. Thermal scattering is, moreover, unlikely to be the only contribution to the reconstructed amplitude. This motivates further investigation into the influence of detector modulation transfer and partial coherence on quantitative amplitude recovery. Understanding these non-thermal contributions will be important for establishing the extent to which the reconstructed amplitude can be interpreted quantitatively.


\section{Competing interests}
No competing interest is declared.

\section{Author contributions statement}

All authors contributed to the development of the research project. B.R.D. performed the simulations and reconstructions and analyzed the results. B.R.D. wrote the manuscript with support from C.G. and J.M.L., and all authors reviewed the manuscript and have given their final approval on the manuscript.

\section{Acknowledgments}

We thank Prof.~Scott Findlay for input on the limitations of the conventional absorptive potential model. We acknowledge support from the Air Force Office of Scientific Research (FA9550-23-1-0667). B.R.D. acknowledges support of the National Science Foundation Graduate Research Fellowship  (Grant No. 2141064). The authors acknowledge the MIT SuperCloud and Lincoln Laboratory Supercomputing Center for providing HPC resources that have contributed to the research results reported within this paper.

\bibliographystyle{unsrtnat}
\bibliography{references}
\end{document}